# Handy Annotations within Oracle 10g

Mrs.Dhanamma Jagli, Ms.Priyanka Gaikwad, Ms.Shubhangi Gunjal, Mr.Chaitanya Bilaware,

[1] Assistant Professor, Department of MCA,
V.E.S. Institute of Technology, Mumbai-74, India.
[2,3,4] M.C.A II nd Year Students
[1]nakrekanti@gmail.com,
[2]gaikwad.priyanka07@gmail.com,
[3]gshubhu@gmail.com,
[4]chaitanya.bilaware@gmail.com.

**Abstract.** This paper describes practical observations during the Database system Lab. Oracle 10g DBMS is used in the data base system lab and performed SQL queries based many concepts like Data Definition Language Commands (DDL), Data Modification Language Commands ((DML), Views, Integrity Constraints, Aggregate functions, Joins and Abstract type . While performing practical during the lab session, many problems occurred, in order to solve them many text books and websites referred but couldn't obtain expected help from them. Even though by spending much time in the database labs with Oracle 10g, tried in numerous ways, as a final point expected output is achieved. This paper describes annotations which were experimentally proved in the Database lab.

# I. Introduction

Relational database query languages, like SQL, have played an essential role in the development of relational database systems [2].

### a) Relational Database

The relational model is today the primary data model for commercial at processing applications. It attained its primary position because of its simplicity, which eases the job of the programmer, compared to earlier data models such as the network model or the hierarchical model. A relational database consists of a collection of tables, each of which is assigned a unique name. Thus, in the relational model the term relation is used to refer to a table, while the term tuple is used to refer to a row. Similarly, the term attribute refers to a column of a table[1][2].

### b) Importance of SQL for relational Model

There are a number of database query languages in use, either commercially or experimentally. In this Paper, the most widely used query language, SQL with Oracle 10g is used. Although it refer to the SQL language as a "query language," it can do much more than just query a database. It can define the structure of the data, modify data in the database, and specify security constraints[2].

# II. About oracle 10g

The Oracle Database (commonly referred to as Oracle RDBMS or simply as Oracle) is an *object-relational database management System* (ORDBMS) produced and marketed by *Oracle Corporation* [3].Oracle10g is the latest version of the Oracle DBMS, released early 2004.One of the main focuses of this release was self management. Central self-diagnostic engine built into core database (Automatic Database Diagnostic Monitor or ADDM) for self managing purpose, it can be seen in below Fig.1.

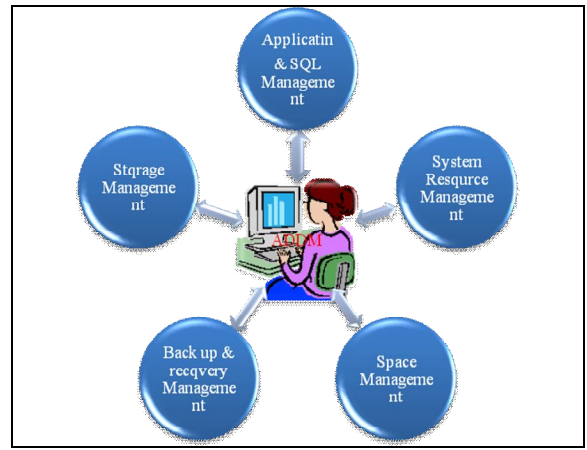

Figure 1: Manageability Infrastructure

### a) History

When Oracle was founded in 1977 as Software Development Laboratories by Larry Ellison, Bob Miner, and Ed Oats, there were no commercial relational database products [2]. The company, which was later renamed Oracle, set out to build a relational database management system as a commercial product, and became a pioneer of the RDBMS market and has held a leading position in this market ever since. Over the years, its product and service offerings have grown beyond the relational database server to include middleware and applications. In addition to products developed inside the company, Oracle's offerings include software that was originally developed in companies that Oracle acquired. Oracle's acquisitions have ranged from small companies to large, publicly traded ones, including People soft, Siebel, Hyperion, and BEA. As a result of these acquisitions, Oracle has a very broad portfolio of enterprise software products. New versions of the products are being developed continually, so all product descriptions are subject to change. The feature set described here is based on the first release of Oracle11g, which is Oracle's flagship database product.

### b) Competition

In the market for relational databases, Oracle Database competes against commercial products such as IBM's DB2 UDB and Microsoft SQL Server. Oracle and IBM tend to battle for the mid-range database market on UNIX and Linux platforms, while Microsoft dominates the mid-range database market on Microsoft Windows platforms. However, since they share many of the same customers, Oracle and IBM tend to support each other's products in many middleware and application categories (for example: WebSphere, PeopleSoft, and Siebel CRM), and IBM's hardware divisions work closely with Oracle on performance-optimizing server-technologies. The two companies have a relationship perhaps best described as "coopetition". Niche commercial competitors include Teradata (in data warehousing and business intelligence), Software AG's ADABAS,Sybase, and IBM's Informix, among many others.

Increasingly, the Oracle database products compete against such open-source software relational database systems as PostgreSQL,Firebird, and MySQL. Oracle acquired Innobase, supplier of the InnoDB codebase to MySQL, in part to compete better against open source alternatives, and acquired Sun Microsystems, owner of MySQL, in 2010. Database products licensed as open source are, by the legal terms of the Open Source Definition, free to distribute and free of royalty or other licensing fees [3].

### c) Types of Oracle 10g Products

**Enterprise Edition** As of July 2010, the database that costs the most per machine-processor among Oracle database editions. The term "per processor" for Enterprise Edition is defined with respect to physical cores and a processor core multiplier. e.g. an 8-processor, 32-core server using Intel Xeon 56XX CPUs would require 16 processor licenses[3].

**Standard Edition** Cheaper, it can run on up to four processors but has fewer features than Enterprise Edition—it lacks proper parallelization, etc.; but remains quite suitable for running medium-sized applications.

**Standard ONE** Sells even more cheaply, but remains limited to two CPUs. Standard Edition ONE sells on a per-seat basis with a five-user minimum. Oracle Corporation usually sells the licenses with an extra cost for support and upgrades which customers need to renew annually.

**Oracle Express Edition (Oracle XE)** An addition to the Oracle database product family offers a free version of the Oracle RDBMS, but one limited to 11 GB of user data and to 1 GB of memory used by the database (SGA+PGA). XE will use no more than one CPU and lacks an internal JVM. XE runs only on Windows and on Linux.

### d) Advanced object-relational Features

Oracle has extensive support for object-relational constructs, including:

- ✓ **Object types.** A single-inheritance model is supported for type hierarchies.
- ✓ **Collection types.** Oracle supports varrays, which are variable length arrays, and nested tables.
- ✓ Object tables. These are used to store objects while providing a relational view of the attributes of the objects.
- ✓ **Table functions**. These are functions that produce sets of rows as output, and can be used in the from clause of a query. Table functions in Oracle can be nested. If a table function is used to express some form of data transformation, nesting multiple functions allows multiple transformations to be expressed in a single statement.
- ✓ **Object views.** These provide a virtual object table view of data stored in a regular relational table. They allow data to be accessed or viewed in an object-oriented style even if the data are really stored in a traditional relational format.
- ✓ **Methods.** These can be written in PL/SQL, Java, or C.
- ✓ **User-defined aggregate functions.** These can be used in SQL statements in the same way as built-in functions such as sum and count.

## III. Annotations

### A. DDL:

Data-definition language (DDL). The SQL DDL provides commands for defining relation schemas, deleting relations, and modifying relation schemas. In this paper, students schema is used for queries purpose. Creating students table:

```
SQL>    CREATE    TABLE    STUDENTS(S_ROLL NUMBER(2),
  2  S_NAME VARCHAR(20),
  3  S_ADDRESS VARCHAR(20),
  4  S_PHONE NUMBER(10),
  5  DOB DATE,
  6  S_MARKS INTEGER);
Table created.
```

After creating table with attribute data type for integer takes default size as 38 and varchar is displaying as varchar2 because of oracle standard, that observation can be seen in the following code.

Describing structure of created table:

```
SQL> DESC STUDENTS;
 Name                            Null?    Type
 ------------------------------- -------- ----------------------------
 S_ROLL                                   NUMBER(2)
 S_NAME                                   VARCHAR2(20)
 S_ADDRESS                                VARCHAR2(20)
 S_PHONE                                  NUMBER(10)
 DOB                                      DATE
 S_MARKS                                  NUMBER(38)
```

### B. DML :

Data-manipulation language (DML). The DML provides the ability to query information from the database and to insert tuples into, delete tuples from, and modify tuples in the database. Inserting records into table in three ways shown in the following.

```
SQL> INSERT INTO STUDENTS
VALUES(1,'ROHI','CHEMBUR',9987918773,'29-
AUG-90',87);
1 row created.

SQL> INSERT INTO
STUDENTS(S_ROLL,S_NAME,DOB,S_MARKS)
VALUES(2,'JUHI','14-APR-91',78);
1 row created.

SQL>   INSERT INTO STUDENTS VALUES
2(&S_ROLL,'&S_NAME','&S_ADDRESS',&S_PHONE
,'&DATE',&S_MARKS);
```

While inserting records into tables, null values are allowed(not for 'not null' attributes)but attribute sequence is required for valid insertion. If attribute is in the middle of the sequence of attributes then it can left blank but if attribute is the last one, then *null* value must be entered as shown in the below code.

```
SQL> INSERT INTO COURSE VALUES
(4,'INFT',);
INSERT INTO COURSE VALUES (4,'INFT',)*
ERROR at line 1:
ORA-00936: missing expression

SQL> INSERT INTO COURSE VALUES (4,'INFT',
NULL)
1 row created.
```

### C. Integrity Constraints:

The DDL includes commands for specifying integrity constraints that the data stored in the database must satisfy. Updates that violate integrity constraints are disallowed. Primary key must be one to any table and it can be used while creating table or can add to existing table also as shown in the following code.

```
SQL> ALTER TABLE STUDENTS ADD CONSTRAINT
  2 PK_P PRIMARY KEY (S_ROLL);
```
After adding primary key by altering table that can be seen in table description as not null.
```
SQL> DESC STUDENTS;
 Name                    Null?      Type
 ------------------      ---------- --------
 S_ROLL                  NOT NULL   NUMBER(2)
 S_NAME                             VARCHAR2(20)
 S_ADDRESS                          VARCHAR2(20)
 S_PHONE                            NUMBER(10)
 DOB                                DATE
 S_MARKS                            NUMBER(38)
Table altered.
```

But whenever other integrity like unique key, check constraints are added by using alter table, reflection can't be seen in table structure as shown in the following.

```
SQL> ALTER TABLE STUDENTS
  2 ADD CONSTRAINT
  3 UK_S UNIQUE(S_PHONE);
Table altered.

SQL>  ALTER TABLE STUDENTS
  2  ADD CONSTRAINT
  3  CHK_S CHECK(S_MARKS >=60);
Table altered.

SQL> DESC STUDENTS;
 Name                    Null?      Type
 ------------------      ---------- --------
 S_ROLL                  NOT NULL   NUMBER(2)
 S_NAME                             VARCHAR2(20)
 S_ADDRESS                          VARCHAR2(20)
 S_PHONE                            NUMBER(10)
 DOB                                DATE
 S_MARKS                            NUMBER(38)
```

If same alter command is used for adding any not null constraint, it can't be added simply using alter. It possible to add not null constraint by using modify command with alter table, which is not similar to other constraints.

```
SQL>  ALTER TABLE STUDENTS
  2  ADD CONSTRAINT
  3* NN DOB NOTNULL;
NN DOB NOTNULL
   *
ERROR at line 3:
ORA-01430: column being added already exists in table

SQL> ALTER TABLE STUDENTS MODIFY DOB DATE
  NOT NULL;
Table altered.
SQL> DESC STUDENTS;
 Name                    Null?      Type
 ------------------      --------   --------
 S_ROLL                  NOT NULL   NUMBER(2)
 S_NAME                             VARCHAR2(20)
 S_ADDRESS                          VARCHAR2(20)
 S_PHONE                            NUMBER(10)
 DOB                     NOT NULL   DATE
 S_MARKS                            NUMBER(38)
```

If someone wants to know the user given constraints for particular table, following syntax can be used but as it was written in oracle user guide, sqlplus is not a case sensitive but here it requires only capital letters in the table name. This observation had shown in the following code.

```
SQL>  SELECT CONSTRAINT_NAME FROM
  USER_CONSTRAINTS WHERE
  TABLE_NAME='STUDENTS';
CONSTRAINT_NAME
-------------------------------
PK_P
UK_S
CHK_S
```

International Journal of Scientific & Engineering Research Volume 4, Issue 1, January-2013
ISSN 2229-5518
5

```
SYS_C005545

SQL>SELECT         CONSTRAINT_NAME         FROM
  USER_CONSTRAINTS                         WHERE
  TABLE_NAME='students';
no rows selected
```

### D. Aggregate functions:

Aggregate functions are functions that take a collection (a set or multiset) of values as input and return a single value[2]. Oracle/SQL offers five built-in aggregate functions: *avg, min, max, sum, count*. With these aggregate functions used on numeric attributes, even name of the attributes is changed in the SELECT query by using key word AS. but with out using that also attribute is renamed as shown in the following code.

```
SQL> SELECT COUNT(S_MARKS) AS TOAL_MARKS
FROM STUDENTS;
  TOAL_MARKS
  ----------
       7
SQL> SELECT  COUNT(S_MARKS)  TOAL_MARKS
FROM STUDENTS;
  TOAL_MARKS
  ----------
       7
SQL>SELECT        COUNT(S_MARKS)        AS
'TOAL_MARKS' FROM STUDENTS;
  SELECT COUNT(S_MARKS) AS 'TOAL_MARKS'
FROM STUDENTS
                *
ERROR at line 1:
ORA-00923: FROM keyword not found where expected

SQL>SELECT         COUNT(S_MARKS)        AS
"TOAL_MARKS" FROM STUDENTS;
  TOAL_MARKS
  ----------
       7
```

### E. Join Operations:

Oracle/SQL provides other forms of the join operation, including the ability to specify an explicit join predicate, and the ability to include in the result tuples that are excluded by natural join[5]. Joins are mainly two types, inner joins and outer joins. Inner join is obtained in two ways as shown in the following Fig.

Figure 2:Inner Join operation Output

## IV. Conclusion

Handy Annotations presented in this paper will be helpful to students and faculties in education sectors and others who wants to use Oracle 10g data base as back end in their applications.